\pdfoutput=1
\documentclass[fleqn,10pt]{wlscirep}
\usepackage[utf8]{inputenc}
\usepackage[T1]{fontenc}

\usepackage{graphicx} 
\usepackage{graphicx,lineno,bm,float,soul,amsfonts,amsmath,siunitx}
\usepackage{xcolor,hyperref}
\usepackage{empheq}
\usepackage{multirow}

\newcommand{\B}[1]{{\bm #1}}

\newcommand{\dd}{\; \text{d}}
\newcommand{\ds}{\displaystyle}

\newcommand{\DVb}{DV_b}
\newcommand{\DVg}{DV_g}
\newcommand{\DE}{DE}

\title{Tangent Velocity constraint for orbital maneuvers with Theory of Functional Connections}

\author[1,*]{A. K. de Almeida Jr.}
\author[1,2]{T. Vaillant}
\author[2,3]{V. M. de Oliveira}
\author[4]{D. Barbosa}
\author[1]{D. Maia}
\author[5,6]{S. Aljbaae}
\author[7]{B. Coelho}
\author[7]{M. Bergano}
\author[4,8]{J. Pandeirada}
\author[5]{A.F.B.A. Prado}
\author[9]{A. Guerman}
\author[2,10]{A.C.M. Correia}
\affil[1]{CICGE, Faculdade de Ciências da Universidade do Porto, 4169-007 Porto, Portugal}
\affil[2]{CFisUC, Departamento de Física, Universidade de Coimbra, 3004-516 Coimbra, Portugal}
\affil[3]{Instituto de Matemática e Estatística, Universidade de São Paulo, 05508-090 São Paulo/SP,Brazil}
\affil[4]{Instituto de Telecomunica\c{c}\~oes, Universidade de Aveiro, 3810-193 Aveiro, Portugal}
\affil[5]{Postgraduate Division - National Institute for Space Research (INPE), São Paulo, Brazil}
\affil[6]{Make The Way, R. Elvira Ferraz, 250 - FL Office 305 e 306 - Vila Olímpia, São Paulo-SP, 04545-015}
\affil[7]{ATLAR Innovation, Ed. Multiusos, Rua Rangel de Lima, 3320-229 Pampilhosa da Serra, Portugal.}
\affil[8]{Instituto Superior Técnico, Avenida Rovisco Pais 1, 1049-001 Lisboa Portugal}
\affil[9]{University of Beira Interior, Covilhã, Portugal}
\affil[10]{IMCCE, UMR8028 CNRS, Observatoire de Paris, PSL Université, 77 Avenue Denfert-Rochereau, 75014 Paris, France}

\affil[*]{Corresponding author: allan.junior@inpe.br}

\begin{abstract}
Maneuvering a spacecraft in the cislunar space is a complex problem, since it is highly perturbed by the gravitational influence of both the Earth and the Moon, and possibly also the Sun. Trajectories minimizing the needed fuel are generally preferred in order to decrease the mass of the payload. A classical method to constrain maneuvers is mathematically modelling them using the Two Point Boundary Value Problem (TPBVP), defining spacecraft positions at the start and end of the trajectory. Solutions to this problem can then be obtained with optimization techniques like the nonlinear least squares conjugated with the Theory of Functional Connections (TFC) to embed the constraints, which recently became an effective method for deducing orbit transfers. In this paper, we propose a tangential velocity (TV) type of constraints to design orbital maneuvers. We show that the technique presented in this paper can be used to transfer a spacecraft (e.g. from the Earth to the Moon) and perform rendezvous maneuvers (e.g. a swing-by with the Moon). In comparison with the TPBVP, solving the TV constraints via TFC offers several advantages, leading to a significant reduction in computational time. Hence, it proves to be an efficient technique to design these maneuvers.
\end{abstract}
\begin{document}

\flushbottom
\maketitle
%
%
\thispagestyle{empty}



\section{Introduction}

In order to perform planetary or interplanetary maneuvers within the solar system, a spacecraft changes its velocity.
Such a variation can be obtained, for instances, with a gravity assist maneuver after a close encounter with a massive body or via specific impulses due to the burn of propellant preloaded in the spacecraft.
In the case of a spacecraft initially orbiting the Earth, a velocity impulse can be applied in a specific direction to guide the spacecraft toward its target.
This is characterized by the flight-path angle, which is the angle between the local horizontal (the line perpendicular to the radius vector) and the velocity vector \cite{vallado7}.
When this angle is equal to zero, the velocity is tangent to the local orbit.
A tangent velocity represents the direction of the initial (or final) velocity after (or before) the application of the impulse of a trajectory that minimizes the costs to transfer a spacecraft between circular orbits\cite{vallado7}.
This result also holds for elliptical orbits, with the point of burn located at either its perigee or apogee \cite{LAWDEN1962323}. A tangential velocity is the solution of the optimization procedure within a two-body problem approach \cite{MARCHAL196991}, although different solutions can take advantage of symmetries for some specific transfers between co-planar elliptical orbits. A tangential initial velocity with respect to the point of burn at the initial circular orbit represents the maximum efficiency to transfer a spacecraft from the Earth to the Moon \cite{Pernicka1995}. This tangential velocity assumption is used to solve the transfer problem (and assess costs) in various works, such as \cite{topputo2013optimal, QI2017106}, employing the shooting method.

The Theory of Functional Connections (TFC) is a recently developed mathematical framework to perform linear interpolation \cite{U-ToC,TFC_Book}.
TFC derives functionals that contain a free function and satisfy linear constraints, regardless of the form of the free function. These functionals enable the search for solutions in a subspace of functions that analytically satisfy the linear constraints. TFC can be applied to transform a constrained problem into an unconstrained one, thereby converting differential equations subject to constraints into unconstrained ones \cite{M-TFC}. For example, these differential equations can represent the equations of motion of a spacecraft \cite{FURFARO202092}. The procedure searches for a solution (finds the trajectory) in a subspace that exactly satisfies the specified linear constraints. This is important, because TFC embeds the type of constraints proposed in this work into the dynamical model. A nonlinear least squares method \cite{MORTARI2019293} is then employed to enhance efficiency of a numerical procedure to solve the problem, and, hence, to design the proposed orbital maneuvers.

The Two Point Boundary Value Problem (TPBVP) in an astrodynamics context consists of searching for a solution that enables a spacecraft to travel from point A to point B. Thus, only the positions of the spacecraft are given, and the velocities at these points are unknown and need to be determined.
One technique for solving this problem is the shooting method, which reduces the problem to a sequence of initial value problems.
The method relies on the integration of the initial conditions multiple times to match the boundary conditions. In astrodynamics, this procedure can leverage solutions of the Lambert's problem as an initial guess \cite{pradobroucke}, can be combined with a direct transcription \cite{direct_transcription} and multiple
shooting approach \cite{topputo2013optimal}, among others in the literature.
The TPBVP can also be efficiently solved using TFC \cite{FURFARO202092}. 
For instance, spacecraft transfers from Earth to the Moon \cite{fastTFC} or between other bodies in the Solar system \cite{allansr} have been evaluated using TFC via the TPBVP constraints.
In this paper, it is proposed the use of a set of constraints different from the TPBVP, called TV (tangential velocity) type constraints. In these constraints, the final position is free to be anywhere on a circumference around the Moon, rather than a single specific point. The approach combines analytical and numerical techniques to efficiently search for solutions satisfying the TV type constraints.

The primary objective of this paper is to introduce the TV type constraints to design orbital maneuvers and demonstrate their efficiency - via the TFC procedure - to transfer a spacecraft from Earth to the Moon and to perform a rendezvous maneuver with the Moon.
The TFC allows us to improve efficiency by analytically embedding the TV type constraints into the equations of motion of the spacecraft to be maneuvered. The proposed method is compared to the TPBVP types of solution also solved via a TFC procedure.
Although the TV type constraints are coupled and nonlinear in rectangular coordinates, as will be shown later, these constraints can become linear after a proper change of coordinates, as was demonstrated in \cite{tfcvariables}. Thus, a set of polar coordinates centered at the Moon is adopted in this paper in order to linearize the proposed constraints.


In Sect. \ref{sec:matdef}, the reference frames and the proper (polar) variables used in this work are defined and the equations of motion are derived.
Sect. \ref{sec:tv} introduces the TV type constraints proposed in this paper, outlines their application scope with TFC, and highlights their advantages in comparison to the TPBVP type of constraints.
In Sect. \ref{sec:transfer}, a practical application of the TV type constraints is provided by transferring a spacecraft from a circular orbit around Earth to a circular orbit around the Moon, aiming to determine the minimum $\Delta V$ required for this maneuver. The advantages of using TV type constraints are also discussed for this maneuver.
Sect. \ref{sec:assist} offers a second application of the TV type constraints involving a gravity assist with the Moon. The influence of the parameters on the efficiency of the maneuver is investigated.
Finally, Sect.~\ref{sec:conclusion} summarizes the main conclusions drawn from this work.

\section{Dynamics modelling}
\label{sec:matdef}

\begin{figure}[t]
    \centering
    \includegraphics[scale=0.4]{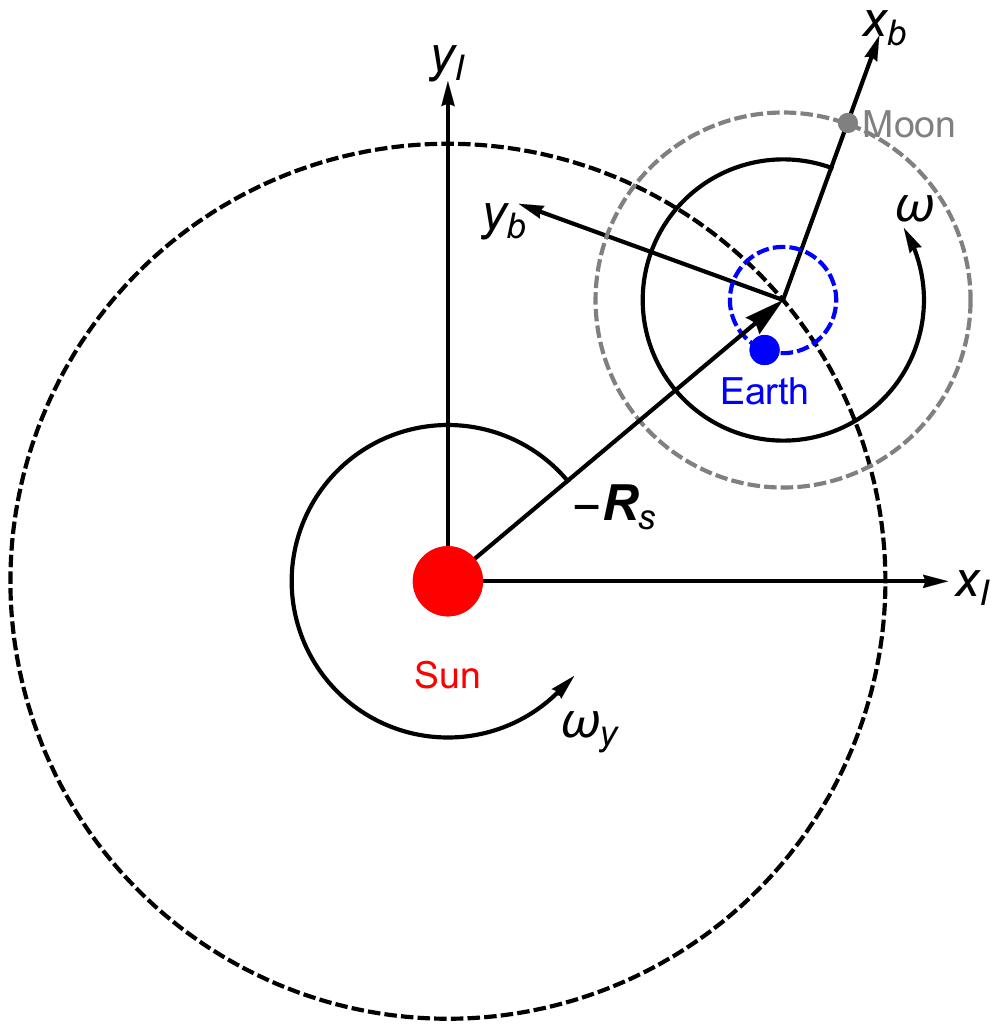}
    \caption{The rotating frame of reference ($x_b,y_b$), with the Earth and the Moon, with respect to the inertial frame of reference ($x_I,y_I$) centered at the Sun.}
    \label{fig:frame0}
\end{figure}

\begin{figure}[t]
    \centering
    \includegraphics[scale=0.4]{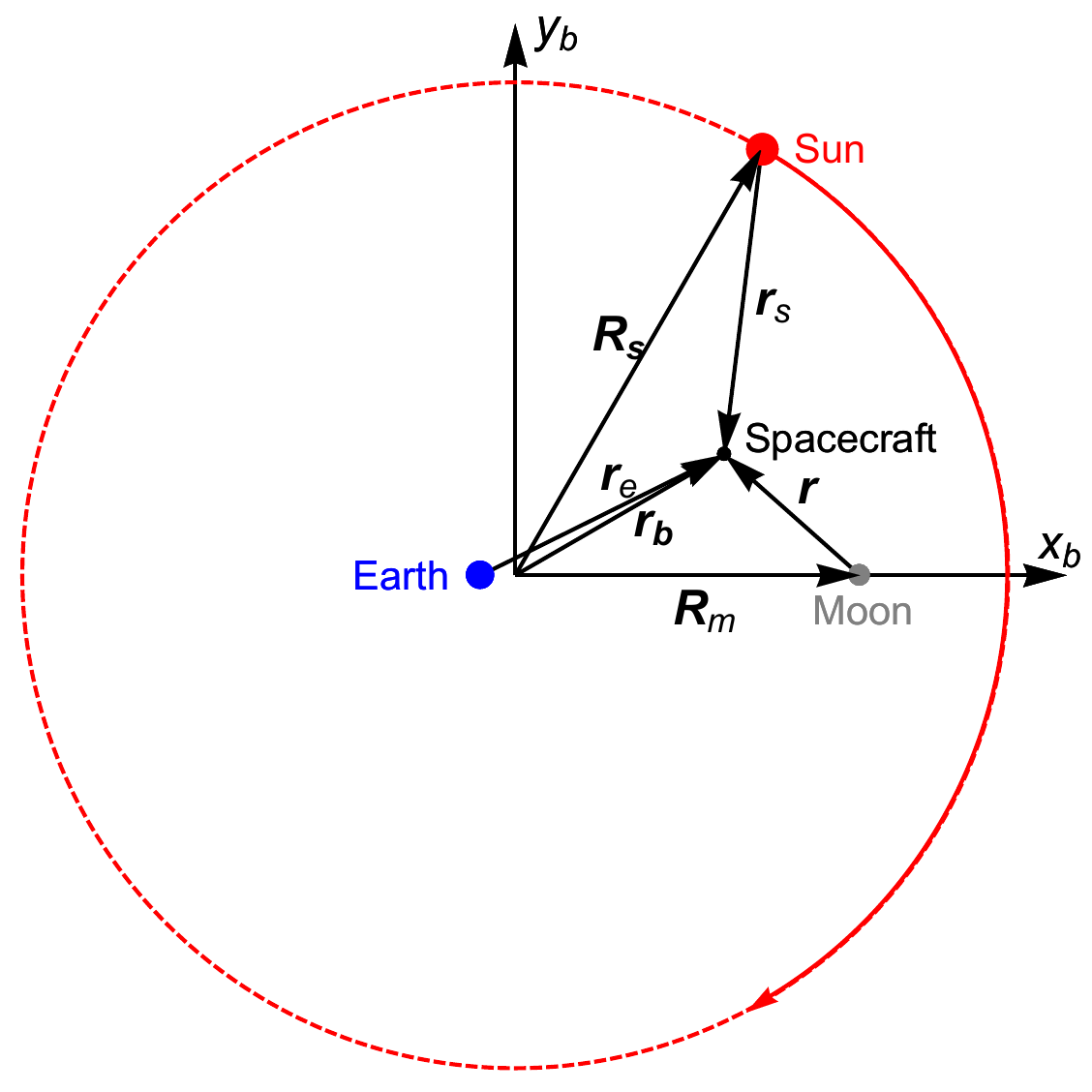}
    \caption{The four bodies and main vectors in the rotating frame perspective.}
    \label{fig:frame1}
\end{figure}

An inertial frame of reference with coordinates ($x_I,y_I$) centered at the Sun is defined.
The Earth and the Moon rotate in circular orbits around their common barycenter, which is the center of a rotating frame of reference with coordinates ($x_b,y_b$) at a distance $R_s$ from the Sun, where $\B{R}_s$ is the position of the Sun with respect to the barycenter. This frame rotates with the same constant angular speed $\omega$ of the Earth and the Moon with respect to the inertial frame. Additionally, its center undergoes rotation around the Sun.
Both frames are co-planar, as shown in Fig.~\ref{fig:frame0} with respect to the inertial frame.
The equations of motion of a satellite in the rotating frame are given by
\begin{equation}\label{eq:4bp}
\begin{aligned}
        	\frac{\dd^2 \B{r}_b}{\dd t^2} + 2 \, \B{\omega} \times \frac{\dd \B{r}_b}{\dd t} + \B{\omega} \times \left(\B{\omega} \times \B{r}_b\right)  =&- \dfrac{\mu_e}{r_e^3} \, \B{r_e} - \dfrac{\mu_m}{r^3} \, \B{r} 
                        - \frac{\mu_s}{r_s^3} \B{r}_s - \frac{\mu_s}{R_s^3} \B{R}_s,
\end{aligned}
\end{equation}
where $\B{r}_b$ is the position of the body with respect to the barycenter in the rotating frame; $\B{\omega} = (0,0,\omega)$ is the angular velocity of the rotating frame; $\mu_e$, $\mu_m$, and $\mu_s$ are the gravitational parameters of the Earth, Moon, and Sun, respectively; $\B{r}_e$, $\B{r}$, and $\B{r}_s$ are the position of the satellite with respect to the Earth, Moon, and Sun, respectively; and $\B{R}_s=\B{r}_b-\B{r}_s$ is the position of the Sun in the rotating frame of reference. The vectors and bodies involved in the motion can be seen in Fig.~\ref{fig:frame1} from the perspective of the rotating frame of reference.

The motion of the satellite must be described relative to the Moon in order to satisfy linearity conditions in the constraints. The position of the satellite can be written as a function of the position relative to the Moon using the transformation
\begin{eqnarray}\label{eq:rrm}
\B{r}_b=\B{r}+\B{R}_m,
\end{eqnarray}
where $\B{R}_m=d_2\hat{\B{x}}_b$ is the position of the Moon relative to the barycenter, with $d_2=R \mu_e/(\mu_e+\mu_m)$, where $R$ is the distance between the Earth and the Moon.
Thus, the equations of motion with respect to the Moon can be derived by replacing Eq.~(\ref{eq:rrm}) into Eq.~(\ref{eq:4bp}), resulting in
\begin{equation}\label{eq:4bpm}
\begin{aligned}
	\frac{\dd^2 \B{r}}{\dd t^2} + 2 \, \B{\omega} \times \frac{\dd \B{r}}{\dd t} + \B{\omega} \times \left(\B{\omega} \times \B{r}\right) - \omega^2 \B{R}_m = &- \dfrac{\mu_e}{r_e^3} \, \B{r_e} - \dfrac{\mu_m}{r^3} \, \B{r} 
                        - \frac{\mu_s}{r_s^3} \B{r}_s - \frac{\mu_s}{R_s^3} \B{R}_s.
\end{aligned}
\end{equation}

\begin{figure}[t]
    \centering
    \includegraphics[scale=0.4]{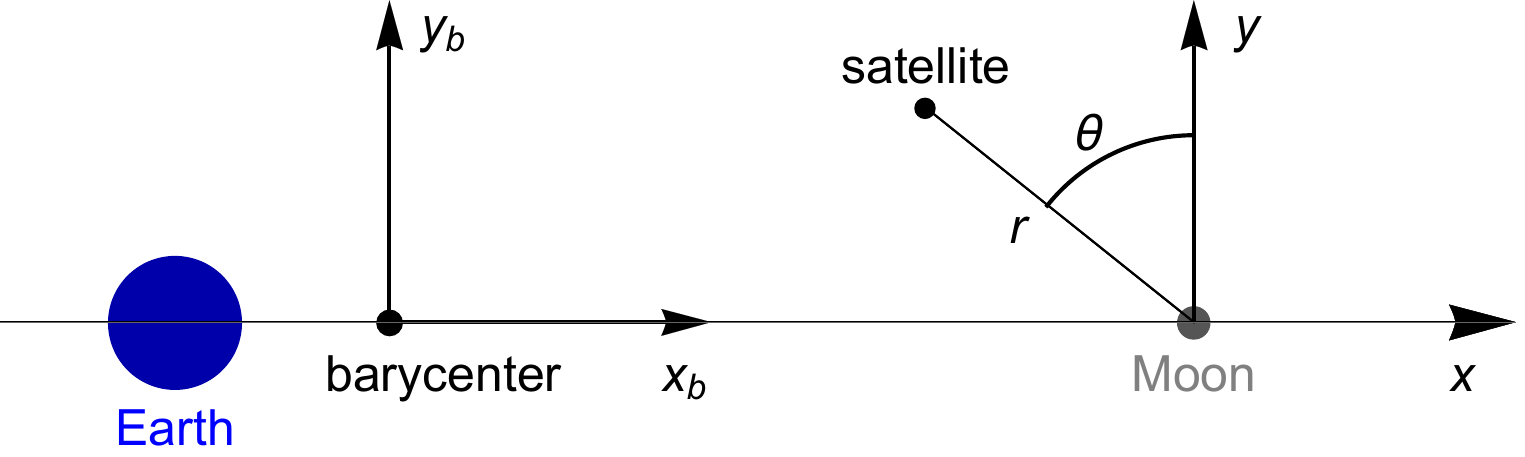}
    \caption{The systems of coordinates rectangular ($x_b,y_b$) centered at the barycenter, rectangular ($x,y$) centered at the Moon, and polar ($r,\theta$) centered at the Moon.}
    \label{fig:polar}
\end{figure}

The components of $\B{r}$ can be written in rectangular coordinates as the pair ($x,y$), defining a system of coordinates centered at the Moon, as illustrated in Fig.~\ref{fig:polar}. Here, $\B{r}$ represents the position of the satellite with respect to the Moon.
This clarification is essential, because the forthcoming constraints will be linearly defined with respect to the Moon, not the barycenter.
The coordinate $r$ is defined as the distance between the satellite and the Moon, while the coordinate $\theta$ is the angle between the position vector $\B{r}$ and the $y$-axis, with positive values in the counterclockwise direction (refer to Fig.~\ref{fig:polar}).
Thus, the polar coordinates are defined based in the following transformation from rectangular coordinates
\begin{equation}\label{eq:referencer}
	\begin{cases} \hat{\B{r}} = -\sin\theta \, \hat{\B{x}} + \cos\theta \, \hat{\B{y}}, \\
             \hat{\B{\theta}} = -\cos\theta \, \hat{\B{x}} - \sin\theta \, \hat{\B{y}}.\end{cases}
\end{equation}
Since $r$ and $\theta$ are variables dependent on time, the position, velocity, and acceleration are given by
\begin{equation}\label{eq:pos}
	\begin{cases}
        \B{r} = r \, \hat{\B{r}}, \\ 
		\dot{\B{r}} = \dot{r} \, \hat{\B{r}} + r \, \dot{\theta} \, \hat{\B{\theta}}, \\
		\ddot{\B{r}} = \left(\ddot{r} - r \, \dot{\theta}^2\right) \hat{\B{r}} + \left(\dfrac{1}{r} \dfrac{\dd}{\dd t} \left(r^2 \, \dot{\theta}\right)\right) \hat{\B{\theta}}.
    \end{cases}
\end{equation}
In accordance with the polar coordinates $(r,\theta)$ defined above, the equations of motion of the satellite given by Eq.~(\ref{eq:4bpm}) become
\begin{eqnarray}
\ddot{r} - r \, \dot{\theta}^2 - r \omega^2 - 2 r \, \omega \, \dot{\theta} + \omega^2 d_2 \sin \theta +\dfrac{ \mu_m}{r^2} +  \dfrac{\mu_e(r-R \sin \theta)}{\left(r^2-2 r R \sin\theta +R^2\right)^{3/2}} - P_r =0, \label{eq:movpol}\\ 
	2 \dot{r} (\dot{\theta} + \omega) + r \, \ddot{\theta} + \omega^2 d_2 \cos \theta -\frac{ \mu_e  R \cos \theta}{\left(r^2-2 r R \sin\theta +R^2\right)^{3/2}} - P_\theta = 0, \nonumber
\end{eqnarray}
where the independent variable is $t$ and the dependent variables are $r(t)$ and $\theta(t)$. 
Here, $P_r$ and $P_\theta$ are the components of the perturbation of the Sun on the satellite in, respectively, $\hat{\B{r}}$ and $\hat{\B{\theta}}$ directions, as expressed by
\begin{equation}\label{eq:prpt}
P_r \hat{\B{r}} + P_\theta \hat{\B{\theta}} = - \frac{\mu_s}{r_s^3} \B{r}_s - \frac{\mu_s}{R_s^3} \B{R}_s,
\end{equation}
where
\begin{equation}
\begin{aligned}
\B{r}_s=&\Big( r  + R_s\cos\theta_s\sin\theta - d_2\sin\theta  - R_s\sin\theta_s\cos\theta     \Big)\hat{\B{r}} 
+\Big( R_s\cos\theta_s\cos\theta - d_2\cos\theta + R_s \sin\theta_s \sin\theta  \Big)\hat{\B{\theta}},
\end{aligned}    
\end{equation}
with magnitude $r_s$ and 
\begin{equation}
\begin{aligned}
\B{R}_s=&\Big(  -R_s \cos\theta_s\sin\theta + d_2\sin\theta + R_s\sin\theta_s\cos\theta  \Big)\hat{\B{r}}  
       +\Big(  -R_s \cos\theta_s\cos\theta + d_2\cos\theta - R_s\sin\theta_s\sin\theta \Big)\hat{\B{\theta}},
\end{aligned}    
\end{equation}
where $R_s=1~\text{AU}$ and $\theta_s$ is the angular position of the Sun relative to the rotating frame centered at the barycenter. This angle is given by $\theta_s=\omega_s t+\gamma$, where $\gamma$ and 
$\omega_s$ are the initial angular position and velocity of the Sun with respect to the rotating frame, respectively.
The values of the parameters used in this work are shown in Table \ref{tab:parameters}. 
For comparison purposes, these values are the same as those used in \cite{simo1995book,fastTFC,Yagasaki2004,topputo2013optimal}.
\begin{table}[ht]
	\centering
	\begin{tabular}{cc}
		\hline
		$\mu_e$&$3.975837768911438\times10^{14}~\text{m}^3/\text{s}^2$ \\[0.5ex]
		$\mu_m$&$4.890329364450684\times10^{12}~\text{m}^3/\text{s}^2$ \\[0.5ex]
		$\mu_s $&$ 1.3237395128595653 \times 10^{20}~\text{m}^3/\text{s}^2$ \\[0.5ex] 
        $\omega$&$2.66186135\times 10^{-6}~\text{s}^{-1}$ \\[0.5ex]
        $\omega_s$&$-2.462743433827215\times 10^{-6}~\text{s}^{-1}$ \\[0.5ex]
        $R$&$3.84405000\times10^{8}~\text{m}$\\[0.5ex]
		\hline
	\end{tabular}
 \caption{The values of the parameters used in this research.}
	\label{tab:parameters}
\end{table}

\section{TV type constraints}
\label{sec:tv}

The constraints introduced in this paper dictate that the spacecraft travels from a point A to any location on a circle of radius $r_p$ around the Moon. Additionally, the final velocity must be such that its component in the radial direction is zero. In this section, these constraints are precisely formulated in mathematical terms and integrated into the equations of motion using TFC.
Subsequently, it is demonstrated that these constraints serve to address two distinct problems: assessing the associated costs of an Earth-to-Moon transfer and evaluating the gains (e.g. specific energy) obtained by utilizing the Moon for a gravity assist maneuver, propelling the spacecraft to higher speeds in the trans-lunar space.

\subsection{Definition}

\begin{figure*}
	\centering
    \includegraphics[width=0.7\linewidth]{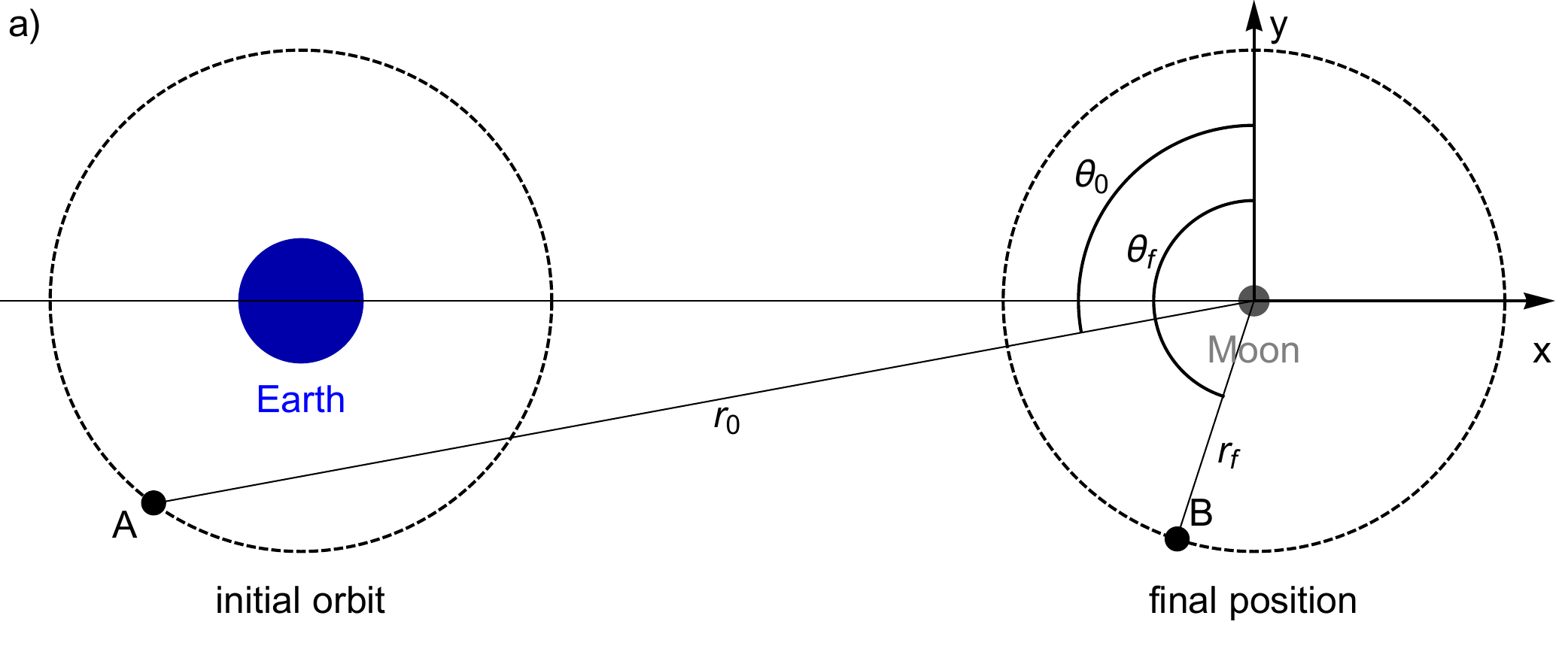}
    \includegraphics[width=0.7\linewidth]{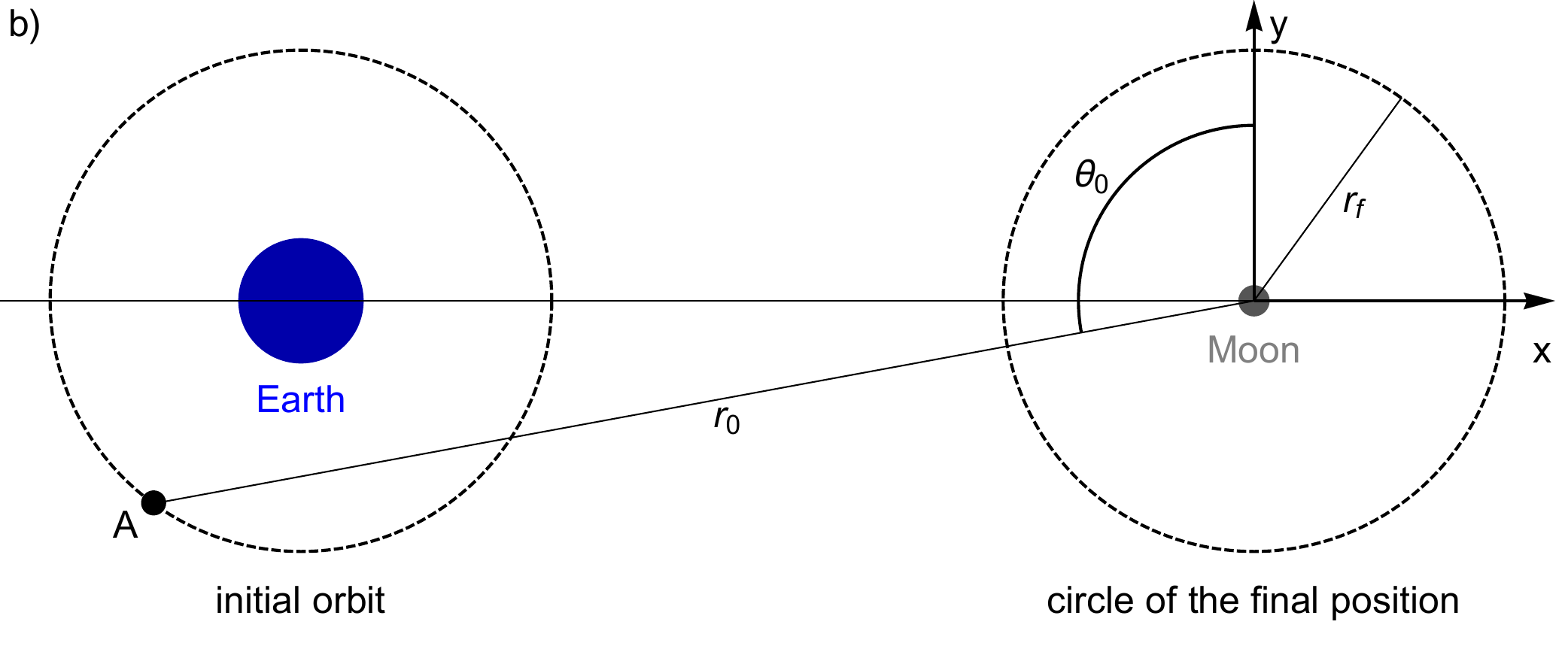}
	\caption{a) The TPBVP involves a transfer from point A to point B without constraints on the velocity. b) The TV type constraints represent a transfer from point A to any location at a distance $r_f$ from the center of the Moon, with tangential velocity relative to the dashed circular final orbit.}
	\label{fig:tv}
\end{figure*}

The TV (Tangential Velocity) type constraints characterize a trajectory starting from an initial point (specified by coordinates $r_0$ and $\theta_0$) at time zero and extending to any point on a circle of radius $r_f$ around the center of the system of coordinates (the center of the Moon) at time $T$.
Furthermore, the velocity at this specified time is such that its radial component is zero.
The mathematical definition of the TV type constraints is shown in Table \ref{tab:constraints}, and a visual representation of these constraints concerning the main celestial bodies is depicted in Fig. \ref{fig:tv}.
It is noteworthy that the final position is not uniquely determined, allowing for any point on a circle around the Moon. This means that, if a solution exists (and a convergence is reached by the method), it will converge to any point at a distance $r_f$ from the center of the Moon such that the other constraint given by a null radial component velocity is also satisfied.
\begin{table}
	\centering
	\begin{tabular}{c|c|c|c|c|c}
		Type&Time&\multicolumn{2}{c|}{Position}&\multicolumn{2}{c}{Velocity}  \\
		& & Radial  & Angular  & Radial  & Angular  \\
		\hline
		\multirow{2}{*}{TPBVP}&$t=0$ & $r = r_0$&$\theta = \theta_0$ &none&none\\
		&$t=T$ & $r = r_f$&$\theta = \theta_f$ &none&none\\
		\hline
		\multirow{2}{*}{TV}&$t=0$ & $r = r_0$&$\theta = \theta_0$ &none&none\\
		&$t=T$ & $r = r_f$& none  &$\dot{r} = 0$&none
	\end{tabular}
	\caption{The constraints of the Two Point Boundary Value Problem (TPBVP) and the Tangential Velocity (TV) proposed in this paper.}
	\label{tab:constraints}
\end{table}

\subsection{Using TFC to embed the TV type constraints into the equations of motion}
\label{sec:embed}

The TFC is extensively developed to address uncoupled and linear constraints on dependent variables \cite{U-ToC}. TFC proves versatile for accommodating any number of constraints of the form
\begin{equation}\label{eq:const1}
\left.\dfrac{\dd^k x}{\dd t^k}\right|_{t_n}=x^{(k)}_n,
\end{equation}
where $k$ is an integer, $t_n$ represents a specific time $t$ value, and $x^{(k)}_n$ denotes the particular value of the $k$-th derivative of $x$ evaluated at time $t_n$. For instance, the combination $n=0$ and $k=0,1$ in Eq.~(\ref{eq:const1}) characterizes the initial value problem (for the time $t_0$), while $n=0,1$ and $k=0$ in Eq.~(\ref{eq:const1}) characterizes the TPBVP (for the times $t_0$ and $t_1$).
TFC can also handle relative constraints of the type
\begin{equation}
\left.\dfrac{\dd^ix}{\dd t^i}\right|_{t_j}=\left.\dfrac{\dd^kx}{\dd t^k}\right|_{t_l},
\end{equation}
where, if $i=k$, then $j\ne l$, and if $j=l$, then $i\ne k$.
Finally, TFC can manage integral type constraints for certain cases of linear equations \cite{mca26030065}.

The TV type constraints at time $T$ represent a fixed final radius around the Moon with a null tangential velocity at this point. In rectangular coordinates, these constraints are expressed as
\begin{empheq}[left={\empheqlbrace}]{align}
    \sqrt{x^2 + y^2}|_{t = T} & = r_f, \\ \nonumber
    (\dot{x} \, x + \dot{y} \, y)|_{t = T} & = 0.
\end{empheq}
In rectangular coordinates, these constraints are coupled and nonlinear. Consequently, applying TFC directly to embed the TV type constraints in these coordinates is not feasible.
However, a proper change of coordinates can be implemented to transform coupled and nonlinear constraints into uncoupled and linear ones \cite{tfcvariables}.
The coordinates $r,\theta$ defined in Sec. \ref{sec:matdef} are adopted in this paper because the TV type constraints are linear on these variables. At time $T$, they become
\begin{empheq}[left={\empheqlbrace}]{align}
    r|_{t = T} & = r_f, \\ \nonumber
    \dot{r}|_{t = T} & = 0.
\end{empheq}
As a result, the constraints are now uncoupled and linear, making it possible to use TFC
to embed them into the equations of motion, as described next.


The constraints shown in Table \ref{tab:constraints} are linear on the dependent variables, allowing them to be incorporated into the equations of motion (Eqs.~\ref{eq:movpol}) using TFC in polar coordinates. Any solution of the resulting equations exactly satisfies these constraints. Following the methodology in \cite{U-ToC}, the \textit{constrained expression} for the dependent variable $r(t)$ is derived from a linear combination of a free function and a given number of linearly independent support functions, where the number of support functions is equal to the number of constraints. Since there are three constraints for the radial coordinate and its derivatives, the \textit{constrained expression} can be derived from
\begin{equation}\label{eq:CEr}
    r(t) = g_r(t) + \eta_0(t, g (t)) s_0(t) + \eta_1(t, g (t)) s_1(t)+ \eta_2(t, g (t)) s_2(t),
\end{equation}
where $g_r(t)$ is the free function associated to the $r$ coordinate and $\eta_k (t, g (t))$ and $s_k(t)$, for $k=0,1,2$, are coefficients and given support functions, respectively. For the purpose of this work, the coefficients $\eta_k (t, g (t))$ are constants in time, and the chosen support functions are $s_0(t)=1$, $s_1(t)=t$, and $s_2(t)=t^2$. The TV type constraints shown in Table \ref{tab:constraints} for the $r(t)$ coordinates are given by
\begin{empheq}[left={\empheqlbrace}]{align}\label{eq:rconst}
    r(0) & = r_0,  \nonumber \\ 
    r(T) & = r_f, \\ 
    \dot{r}(T) & = 0. \nonumber
\end{empheq}
A set of three equations are generated by substituting $r(t)$ from Eq.~(\ref{eq:CEr}) into Eqs.~(\ref{eq:rconst}).
This set of equations is solved for the three coefficients $\eta_k$ for $k=0,1,2$. This operation ensues the derivation of the following \textit{constrained expression} for $r(t)$
\begin{equation}\label{eq:cer}
\begin{aligned}
            r(t) =&  g_r(t) + (-r_f t (t - 2 T) + r_0 (t - T)^2  
        - (t - T)^2 g_r(0) + t (t - 2 T) g_r(T) + t T (-t + T) g^{\prime}_r(T))/T^2,
\end{aligned}
\end{equation}
where $g_r^\prime(T)$ represents the derivative of the free function $g_r(t)$ with respect to $t$ evaluated at $t=T$.
Analogously, the \textit{constrained expression} for the $\theta$ variable can be derived as
\begin{eqnarray}\label{eq:cetheta}
    \theta(t) = g_\theta(t) + \theta_0 - g_\theta(0),
\end{eqnarray}
where $g_\theta(t)$ is the free function for the $\theta$ coordinate.
The free functions are written as
\begin{equation}\label{eq:ff}
	\begin{cases}
        g_r (t) = \ds\sum_{j = 3}^m \xi_j \, h_j (\tau), \\
        g_\theta (t) = \ds\sum_{j = 1}^m \zeta_j \, h_j (\tau),
    \end{cases}
\end{equation}
where $\xi_j$ and $\zeta_j$ are unknown coefficients, $h_j$ are the orthogonal Chebyshev polynomials (e.g. \cite{abramowitzstegun1972}), and $m$ is its highest order truncation. It is important to note that the orthogonal Chebyshev polynomials are valid only in the interval $-1\le \tau \le 1$, hence the change of the independent variable from $t$ to $\tau$ given by
$\tau=2t/T-1$ 
must be performed
in order to satisfy this time interval. 
Note that the sum starts at $j=3$ and $j=1$ for $g_r$ and $g_\theta$, respectively, because  the terms of the polynomials must be linearly independent with the support functions, and there are three support functions (for the three constraints given by Eqs.~(\ref{eq:rconst})) for $r$ (and its derivative) and one only support function (for the constraint $\theta(0)=\theta_0$) for $\theta$.

The free functions shown in Eqs.~(\ref{eq:ff}) are used in the expressions for the $r$ and $\theta$ variables shown in Eqs.~(\ref{eq:cer}) and (\ref{eq:cetheta}). After this, the dependent variables $r$ and $\theta$ in Eqs. (\ref{eq:movpol}) are replaced by the respective ones given by Eqs.~(\ref{eq:cer}) and (\ref{eq:cetheta}).  This procedure ensues in a set of two unconstrained differential equations. This set is discretized 
through the collocation points method
using the Chebyshev–Gauss-Lobatto nodes \cite{lanczos1988applied} distributed between 0 and $T$ as follows
\begin{equation}\label{eq:distr}
     t_k = \bigg(1- \cos \bigg(\frac{k \pi}{ N} \bigg) \bigg) \frac{T}{2}, \quad \text{for}~k=0,1,2,...,N,
\end{equation}
where $(N+1)$ is the number of equations generated by this procedure for each equation in Eqs.~(\ref{eq:movpol}). These equations are then solved for the unknown coefficients $\xi_j$ for $j=3,4,...,m$ and $\zeta_j$ for $j=1,2,...,m$ using the nonlinear least squares optimization method \cite{M-ToC} to minimize the sum of the squares of the left sides of Eqs.~(\ref{eq:movpol}) under the distribution given by Eq.~(\ref{eq:distr}). Further details on the TFC procedure can be seen in \cite{U-ToC,M-ToC}.

\subsection{Comparison with the TPBVP}
\label{sec:comparison}

The problem represented by the TPBVP is to transfer a spacecraft from point A around the Earth to point B around the Moon in a given transfer time $T$, as seen in Fig. \ref{fig:tv} (a). 
The constraints of the type TPBVP are shown in Table \ref{tab:constraints}. Although two positions are specified at two times, there is no constraint on the velocity.

In this paper, it is proposed the use of the TV type constraints for this problem. These constraints are identical to the traditional TPBVP at the initial time ($t=0$), but at the final time ($t=T$), a constraint is applied in the radial component of the final velocity $\dot{r}(t)|_{t=T}$, instead of applying the constraint at the angle $\theta(T)$.
A comparison between the TV and TPBVP types of constraints is shown in Table \ref{tab:constraints}. 
The main distinction between the TV and TPBVP types is that the final velocity and position are perpendicular in the TV type, i.e. $(\B{r}\cdot\dot{\B{r}})|_{t=T}=0$. Since the system of coordinates is centered at the Moon, the final velocity is tangential to a final circular orbit of radius $r|_{t=T}=r_f$ to which the spacecraft is transferred.


As was mentioned earlier, for maximum efficiency of the burn, tangential burn occurs only at apogee or perigee on elliptical orbits \cite{vallado7,LAWDEN1962323}. This result is valid in a two-body problem local approximation \cite{MARCHAL196991}. In a three body-problem context, the impulse applied in the same direction as the velocity relative to the rotating frame maximizes the change in the Jacobi constant \cite{Pernicka1995}. In a restricted four body-problem context, a tangential impulse is assumed in the numerical procedure given by the shooting method to evaluate the costs for Earth-to-Moon transfers in \cite{topputo2013optimal}, which is adopted also using lunar gravity assist \cite{QI2017106}.
A tangential burn applied at perigee or apogee is represented by the tangential velocity constraint (included in the TV type constraints) and it is analytically integrated into the equations of motions using TFC. Hence, any solution numerically obtained using the procedures explained in this section must satisfy this constraint, which is not present in the TPBVP.
Besides that, the tangential velocity in the two-body context means that this point is the periapsis of the orbit around the Moon, so there is no risk of collision of the spacecraft with the Moon.
In fact, the motion close to the Moon for short periods of time can be approximated by the two-body problem. 
The TV type constraints can be visualized in Fig. \ref{fig:tv} (b) with the dashed circular initial and final orbits of the transfer. Note that the point $B$ is not defined in the final orbit because there is no constraint in the final angle $\theta (T)$.

\section{Earth-to-Moon transfer using the TV type constraints}
\label{sec:transfer}

The task in this section is to transfer a spacecraft from an initial circular orbit around the Earth to another circular orbit around the Moon within a specified time transfer duration $T$ using bi-impulsive maneuvers. A first impulse $\Delta V_1$ is applied at the initial orbit around the Earth to initiate the transfer, and another impulse $\Delta V_2$ is applied at the final time $t=T$ to circularize the orbit around the Moon. The spacecraft is transferred under the gravitational influence of the Earth, Moon, and Sun.
This problem can be formulated as a TPBVP. Several methods are available to solve it, such as the patched restricted three-body problem \cite{da2012optimal,5586384}, gradient shooting method \cite{pradobroucke}, direct transcription and multiple shooting method \cite{topputo2013optimal}, and Jacobi integral variational equation \cite{GAGGFILHO2019312}. 
The bi-circular restricted four-body problem has been commonly employed for such transfers in numerous works
\cite{topputo2013optimal,1992sfm..proc.1113Y,Yagasaki2004,dasilvamar2011,doi:10.2514/1.55426,Oshima2017,Onozaki2017,QI2017106,oshimatop19}. 
Furthermore, the TPBVP is solved using TFC for the Earth-to-Moon transfer problem based on the bi-circular restricted four-body problem in \cite{fastTFC}.
In this paper, an alternative approach is proposed by adopting the TV type constraints outlined in Sect. \ref{sec:tv}, instead of the TPBVP, to address the Earth-to-Moon transfer problem. The efficiency of the solution is enhanced by the TFC procedure described in Sect. \ref{sec:embed}.

\subsection{Advantages of the TV type constraints}

The initial position must be specified for both the TPBVP and the TV types of constraints. On the other hand, the angle of the final position must be provided only when using TPBVP, because the final angle is unconstrained in the TV type constraints. 
The TV type constraints show advantages in comparison with the TPBVP type in evaluating the costs for transfers between the Earth and the Moon. They are:
\begin{itemize}
    \item The value of $\theta (T)$ satisfying $\dot{r}|_{t = T} = 0$ (to minimize costs) is directly provided by the solution (by the convergence of the optimization method) when using TV type constraints.
    On the other side, for the TPBVP constraints, an additional optimization procedure must be implemented in order to find the angle of the position satisfying $\dot{r}|_{t = T} = 0$.
    Hence, the numerical solution provided by using the TV type is faster than TPBVP type. A quantitative comparison will be shown in Sect. \ref{sec:comp}.
    \item There is no need to use patched conics to solve the transfer problem. 
    In patched conics, the motion is cut out in different sections, and each of them corresponds to a two-body approximation, where the main body is the one which has the highest gravitational influence on the spacecraft.
    \item The use of integrators (such as Runge-Kutta, etc.) in the code is unnecessary, as required by shooting methods to solve TPBVP type constraints. There is no need to iterate the initial conditions multiple times to match boundary conditions.
    \item The solution ensures no collision with the Moon. Assuming that the velocity of the satellite in the rotating frame is approximately equal to its relative velocity with the Moon in an inertial frame centered at the Moon (in a two-body problem context, as shown in \cite{Broucke1988}), the condition $\dot{r}|_{t = T} = 0$ implies that $\B{r}(T)$ is locally the periapsis (or apoapsis) of the orbit. In a two-body problem context, the spacecraft has zero radial velocity only at the periapsis or apoapsis. The position $\B{r}(T)$ cannot be the apoapsis, since the spacecraft comes from a region out of the sphere of influence of the Moon.
    \item 
    Since the risk of collision with the Moon is null, there is no need to choose a large $r_f$, contrary to TPBVP.
\end{itemize}

\subsection{Results for a bi-impulsive maneuver to solve the Earth-to-Moon transfer}
\label{sec:num_transfer}


This subsection presents the costs associated to the use of bi-impulsive maneuvers to transfer a spacecraft from an initial circular orbit of 167 km altitude around the Earth to a final orbit of 100 km altitude around the Moon (considering a radius of the Earth and Moon of $6378~\text{km}$ and $1738~\text{km}$, respectively). The total cost $\Delta V=\Delta V_1+\Delta V_2$ is the sum of the impulse $\Delta V_1$ applied at the initial orbit around the Earth and the one applied at the final orbit around the Moon $\Delta V_2$. 
For comparison purposes, these costs are evaluated in this paper using both the TPBVP and TV types of constraints defined in Sect. \ref{sec:tv}. Note that the solutions obtained using these constraints are deduced for a given initial point $A$, which can be anywhere in the initial orbit around the Earth (see point $A$ in Fig. \ref{fig:tv}). Hence, an additional procedure must be implemented in order to find the best point $A$ related to the lowest cost $\Delta V$, as described next:
\begin{itemize}
\item Simulations are launched for each position $(r_0,\theta_0)$ of point A in the initial orbit of 167 km altitude around the Earth (Fig. \ref{fig:tv}) spanned with a first rough step.
\item For each position $(r_0,\theta_0)$, the cost $\Delta V$ is computed from the simulation.
\item The position $(r_0,\theta_0)$ presenting the minimum value of $\Delta V$, denoted as $(r_0,\theta_0)_{min1}$, is determined.
\end{itemize}
This procedure is repeated, but this time in a smaller interval of the orbit centered on $(r_0,\theta_0)_{min1}$ with a smaller spanning step, in order to determine a more accurate position $(r_0,\theta_0)_{min2}$ minimizing the value of $\Delta V$, and so on.
The procedure is then repeated several times by successively decreasing the step until a new iteration of the procedure no longer modifies the position $(r_0,\theta_0)$ minimizing $\Delta V$ within a given accuracy. 
Simultaneously, the same procedure is applied for the initial position of the Sun, indicated by the angle $\gamma$.
The TV type constraints do not require an additional procedure to find the best point $B$ (with the lowest cost $\Delta V$) to apply the second impulse for the final circular orbit.
However, if the TPBVP type is adopted, an additional procedure must also be implemented to find the best point B on the final orbit.

\begin{figure*}
	\centering
    \includegraphics[width=0.6\linewidth]{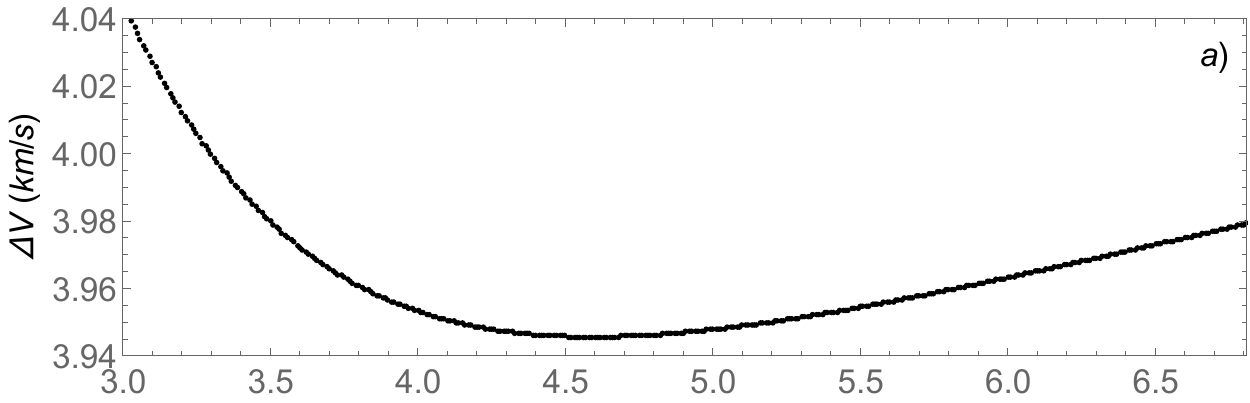}
    \includegraphics[width=0.61\linewidth]{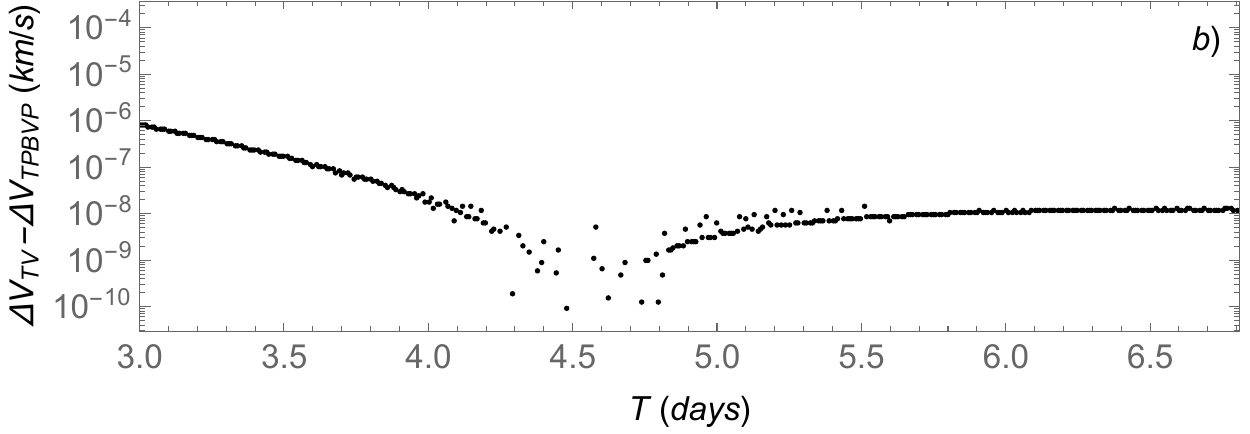}
	\caption{The costs of Earth-to-Moon transfers evaluated using the TV type constraints as function of flight time $T$ are shown in a). The difference between these costs evaluated using the TV and the TPBVP types of constraints is shown in b).}
	\label{fig:costs}
\end{figure*}

The $\Delta V$ costs evaluated using the TV type constraints are depicted in Fig. \ref{fig:costs} (a)
as a function of the flight time $T$, where $3\leq T \leq 6.8$ days with a step of $0.01$ day. These costs consider the gravitational influence of the Sun. The lowest cost found in this range is 3945.6619 m/s, associated with a flight time $T=4.59$ days. This cost aligns with other works in the literature that also consider the solar perturbation for this Earth-to-Moon transfer.
For instance, the minimum cost of $3949.53$ m/s associated with $T=4.59446$ days is shown in \cite{Yagasaki2004}, the minimum cost of $3944.8$ m/s associated with $T=4.6$ days in \cite{topputo2013optimal}, and the minimum cost of $3944.83$ m/s associated with $T=4.625$ days in \cite{fastTFC}. In comparison, a Hohmann-like transfer requires $5$ days with a total cost of $3954$ m/s \cite{topputo2013optimal}.
The same $\Delta V$ costs shown in Fig. \ref{fig:costs} (a) were also evaluated using the TPBVP type constraints, but this time with the aid of the additional procedure to find the position of the point $B$ in the final orbit (the value of $\theta_f$) that minimizes the $\Delta V$.
The differences between the $\Delta V$ evaluated using the TV and TPBVP types of constraints are shown in Fig. \ref{fig:costs} (b).
Note that these differences are very small. The closer to the minimum $\Delta V$, the smaller this difference. Such a difference around $T\approx4.59$ days could not be properly evaluated in this narrow region because it is smaller than the machine level error involved in the numerical procedures.
This is an important result, since the very small differences between the TV and the TPBVP types of constraints shown in Fig. \ref{fig:costs} (b) means that the TV type is almost as accurate as the TPBVP type to minimize the $\Delta V$ costs. 

\subsection{Computational efficiency}
\label{sec:comp}

In this subsection, a comparison between the traditional TPBVP and the proposed TV types of constraints is done concerning the numerical efficiency of the methods.
The Python language with the assistance of the TFC module \cite{tfc2021github} is chosen to numerically evaluate the NLLS method using singular-value decomposition in a just-in-time (JIT) compiler \cite{JaxGithub}.
This procedure is adopted to solve both the TPBVP and TV types of constraints by embedding the respective constraints into the equations of motion (Eqs.~\ref{eq:movpol}).
In order to determine the coefficients $\xi_j$ and $\zeta_j$ of the free functions (Eq. \ref{eq:ff}), the TFC method proceeds iteratively with a nonlinear least squares procedure to best fit the trajectory with the dynamical equations.
An initial guess is then necessary.
The first initial guess adopted is a trajectory with a constant velocity between a point with a distance of 167 km from the surface of the Earth and a point with an altitude of 100 km from the surface of the Moon.
After a first iteration, a new set of coefficients is obtained and it is used as an initial guess for the procedure explained in Sect. \ref{sec:embed}.
The procedure is then solved for a flight time $T=1$ day, and the solution obtained is used as a initial guess for the following flight time, increased by the step used in this work, i.e., 0.01 day.

The average times to find the minimum costs for a given flight time $T$ are shown in Table \ref{tab:effic}. They are evaluated from an ensemble of 100 cases each. The TV type constraints show a reduced computational time of $31.39\%$ in comparison with the TPBVP type of constraint for the 4BP case (considering the perturbation of the Sun), and also a reduced computational time of $45.35\%$ for the 3BP case.
Note that the TFC method to obtain the solution is very fast - a solution is usually obtained in less than one second for a given set of constraints. The major computational cost is the additional optimization procedure, which must be implemented in the code to find the position of the point $A$ of the initial orbit, the initial best position of the Sun (the angle $\gamma$ in the case of the 4BP), and/or the position $B$ of the final orbit around the Moon in the case of the TPBVP. This last optimization procedure (to find the final position around the Moon) is not needed in case the TV type constraints are used. This is the main advantage of the TV type in comparison with the TPBVP type constraints. 

A residual is defined as the sum of the magnitude of the left side of Eqs.~(\ref{eq:movpol}) allocated for the $N+1$ different times $t_k$ given by Eq.~(\ref{eq:distr}). The maximum value of the order of magnitude of the residual divided by N is of the order of $10^{-11}$ (m/s$^2$) for all cases among the TV or TPBVP for the 3BP or 4BP.
\begin{table}
    \centering
    \begin{tabular}{c|c|c}
         Type of&\multicolumn{2}{c}{Computational time (s)}  \\
         constraints&4BP&3BP  \\
          \hline
         TPBVP&49.6575&38.4789 \\
         \hline
         TV&34.0725&21.0278  \\
    \end{tabular}
    \caption{Average time to find the minimum $\Delta V$ cost for a given flight time. The number of points discretized in the time domain and the number of basis functions are such that $N=400$ and $m=396$. The transfer using the TPBVP and the TV types of constraints are both solved using the TFC procedure.}
    \label{tab:effic}
\end{table}



\section{Moon's gravity assist maneuver}
\label{sec:assist}

\begin{figure}[t]
    \centering
    \includegraphics[width=0.6\linewidth]{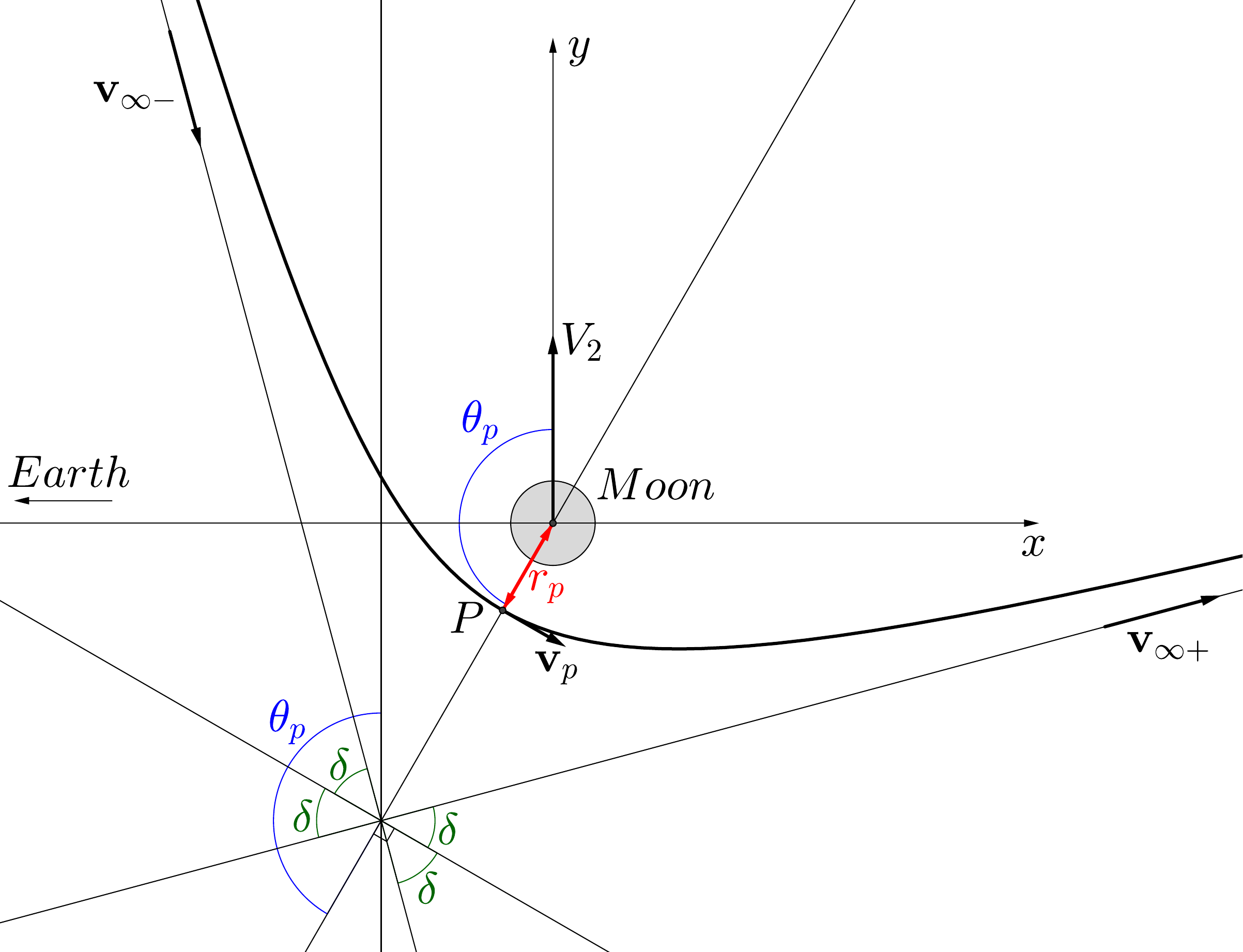}
    \caption{Parameters of a gravity assist with the Moon (following \cite{Broucke1988}). $P$ is the periapsis.}
    \label{fig:hyper}
\end{figure}

A gravity assist maneuver can be completely described by the following three parameters \cite{Broucke1988}, which are also shown in Fig. \ref{fig:hyper}:
\begin{itemize}
    \item the periapsis distance $r_p$,
    \item the magnitude of the velocity at the periapsis $v_p$,  
    \item the close approach angle $\theta_p$.
\end{itemize}
The efficiency of the gravity assist can also be quantified by the velocity gain which corresponds to the increase of velocity of the spacecraft with respect to the Earth-Moon barycenter due to the rendezvous maneuver with the Moon. The gain in velocity is defined as 
\begin{equation}\label{eq:dv1}
\DVb =\| \B{V}_f - \B{V}_i \|,
\end{equation}
where $\B{V}_i$ and $\B{V}_f$ represent the velocities of the spacecraft at the sphere of influence of the Moon in an inertial frame of reference  centered at the barycenter of the Earth-Moon system before and after the rendezvous maneuver, respectively.
Assuming that the motion of the spacecraft close to the Moon can be approximated by the two-body problem dynamics, this gain is \cite{Broucke1988}
\begin{equation}\label{eq:dvb}
    \DVb = 2 v_\infty \sin\delta,
\end{equation}
where the angle $\delta$ (Fig. \ref{fig:hyper}) is defined by
\begin{equation}
    \sin\delta=\frac{1}{1+\frac{r_p v_\infty^2}{\mu_m}},
\end{equation}
and $v_\infty$ is the norm of the relative velocity of the spacecraft with respect to the Moon when approaching it. 
The constant energy of the two-body problem is used to connect this velocity to the velocity at the periapsis, according to \cite{Broucke1988}
\begin{equation}\label{eq:dv3}
     v_\infty^2 = v_p^2 - 2 \frac{\mu_m}{r_p}.
\end{equation}
The gain $\DVb$ shows the change in velocity due to the gravity assist, but cannot indicate if the velocity increases or decreases. 
Hence, another gain in velocity can be defined by
\begin{equation}\label{eq:dvg}
    \DVg = \left\lVert \B{V}_f\right\rVert -\left\lVert \B{V}_i\right\rVert.
\end{equation}
It measures the gain in the norm of the velocity with respect to the Earth, and can be deduced from the expression of the velocity given in \cite{Broucke1988} with
\begin{eqnarray}
\left\lVert \B{V}_f\right\rVert & = & \sqrt{v_\infty^2+V_2^2-2v_\infty V_2\sin\left(\theta_p+\delta\right)}, \label{eq:Vf}\\
\left\lVert \B{V}_i\right\rVert & = & \sqrt{v_\infty^2+V_2^2-2v_\infty V_2\sin\left(\theta_p-\delta\right)},
\end{eqnarray}
where $V_2=d_2 \omega$ is the velocity of the Moon with respect to the barycenter. 
The specific energy gain is given by \cite{Broucke1988}
\begin{equation}\label{eq:dener}
    \DE = -2 V_2 v_\infty \cos \theta_p\sin\delta.
\end{equation}
Note that the notation used by \cite{Broucke1988} to characterize the gravity assist maneuver is $\gamma$, which is the angle between the periapsis and the $x$ axis. In this paper, $\theta_p$ is the angle between the periapsis and the $y$ axis. This choice enables a direct relation with $\theta$ defined in Sect. \ref{sec:matdef} and shown in Fig. \ref{fig:polar} for the case where the position is such that $\theta(T)=\theta_p$. Hence, the relation between the angle adopted in this paper and the angle used in \cite{Broucke1988} is $\theta_p = \gamma - 90^{\circ}$. The relation $\sin \gamma = \cos \theta_p$ is used to write Eq.~(\ref{eq:dener}) obtained from the respective energy gain shown in \cite{Broucke1988}.
Note that from this equation, the specific energy decreases for $-90^{\circ}<\theta_f<90^{\circ}$, and increases for $90^{\circ}<\theta_f<270^{\circ}$, with a maximum at $180^{\circ}$.


\subsection{Using the TV type constraints to perform a gravity assist maneuver}

The first parameter of interest in a gravity assist maneuver is the distance at the perigee $r_p$. In this maneuver, we propose that $r_p$ be given by the final position of the transfer. The connection between the perigee and the TV type constraint is 
\begin{equation}\label{eq:r_p}
    r_p=r_f.
\end{equation}
The maneuver is then evaluated as a function of this distance, which must be provided. The following parameter of the transfer is the magnitude of the velocity at periapsis $v_p$. In a two-body problem, the radial component of the velocity at the periapsis is null, hence, this magnitude is evaluated only with the tangential component $v_p = r_p \dot{\theta}_p$. Since $r_p$ is provided, the parameter of the gravity assist maneuver becomes $\dot{\theta}_p$, instead of $v_p$. The connection between this parameter and the TV type constraints is 
\begin{equation}\label{eq:dottheta}
\dot{\theta}_p = \dot{\theta}_f.
\end{equation}
Hence, the velocity at the periapsis is $v_p = r_f \dot{\theta}_f$.
The last parameter of the gravity assist maneuver is $\theta_p$. The connection between this parameter and the TV type constraints is
\begin{equation}\label{eq:theta}
\theta_p = \theta_f.
\end{equation}


\subsection{Advantages}
\label{sec:advan_GA}

The constraints of the type TV show the following advantages in comparison with the type TPBVP:
\begin{itemize}
    \item In the procedure proposed in this manuscript to perform the gravity assist maneuver, the parameters $\dot{\theta}_p$ and $\theta_p$ are obtained by the solution (through the convergence) of the optimization method.
    \item It is not necessary to use patched conics to solve the transfer problem.
    \item There is no need to use optimization procedures to satisfy the main constraint of the rendezvous maneuver, which is $\dot{r}|_{t = T} = 0$ for a given flight time, because it is embedded into the equations using TFC.
    \item Once again, since $r_f$ is the periapsis ($r_f=r_p$), the solution does not show any collision with the Moon.
    \item Since a collision with the Moon is not possible, there is no need to choose a large $r_f$ (in comparison with methods to solve the TPBVP) in order to avoid solutions that collide with the Moon.
    \item The periapsis distance $r_p$, which is one of the three parameters of the gravity assist, is specified in the constraint.
    It is then easier to study its influence on the trajectory, and to determine what initial conditions must be chosen in order to have a gravity assist with a given value of periapsis distance.
    In the TPBVP, the periapsis distance can only be deduced a posteriori from the obtained solutions as the two other parameters (the magnitude of the velocity at the periapsis $v_p$,  the close approach angle $\theta_p$).
    With the TV type constraints, the periapsis is imposed and the two other parameters are obtained from the solution.
\end{itemize}

\subsection{Methodology}

In the rendezvous maneuver, because it is not necessary to put the spacecraft in an orbit around the Moon, only one impulsive maneuver is applied, contrary to the transfer maneuver where two impulses are used.
The spacecraft starts from a circular orbit at an altitude of 167 km around the Earth, and a unique impulse characterized by $\Delta V$ is then applied so that the spacecraft heads towards the Moon for a gravity assist.
Here, the influence of the Sun is not considered in order to simplify the problem. The terms due to its perturbation are set to zero ($P_r=0$ and $P_\theta=0$) in the equations of motion given by Eqs.~(\ref{eq:movpol}),
and the problem is reduced to the CR3BP (with the inertial frame centered at the Earth-Moon barycenter for simplicity).
The influence of the Sun is relatively weak for a short time (less than 10 days) orbit transfer from the Earth to the Moon. 
For instance, the influence of the Sun can reduce the costs of an Earth-to-Moon transfer by about 5 m/s for short time travels \cite{fastTFC}. Although this value may be significant for a mission, it represents a small percentage of the total cost. Thus, it is expected that the perturbation of the Sun does not change the general behavior of the gain for a gravity assist maneuver, as was noted, for instance, in \cite{negri2019}.

Like in Sect. \ref{sec:num_transfer}, an additional optimization procedure is performed in order to find the most efficient trajectory, where, from the initial impulse $\Delta V$, the final velocity after the gravity assist $V_f=\left\lVert \B{V}_f\right\rVert$ is as high as possible.
Here, the optimization with TFC gives a solution for the trajectory between the starting point A and the periapsis, and then all three parameters of the gravity assist maneuver are known (the periapsis distance $r_p$, the magnitude of the velocity at the periapsis $v_p$, the close approach angle $\theta_p$).
From these parameters, the velocity $V_f$ is then estimated from Eq.~(\ref{eq:Vf}), which was obtained from a two-body approximation.
In order to obtain an efficient gravity assist, the procedure then minimizes $\Delta V/V_f$, which represents the relative velocity gain.
The optimization procedure used for this is the following:
\begin{itemize}
\item The initial circular orbit of 167 km of altitude around the Earth is spanned with a rough first step, and simulations are carried out for each of the initial positions $(r_0,\theta_0)$ obtained.
\item For each position $(r_0,\theta_0)$, the ratio $\Delta V/V_f$ is computed from the simulation.
\item The position $(r_0,\theta_0)$, that provides the lowest value of $\Delta V/V_f$, labeled $(r_0,\theta_0)_{min}$, is then determined.
\end{itemize}
This operation is then repeated several times, as it was done in Sect. \ref{sec:num_transfer}, in order to refine the position $(r_0,\theta_0)_{min}$, always working with smaller intervals around this value and smaller steps.
It is important to note that the specific position of the periapsis represented by $\theta_p$ in Fig. \ref{fig:hyper} is obtained by the convergence of the procedure. 
The use of the TV type constraints allows the code to converge to a value of $\theta_p$ that minimizes the objective function for a given flight time. This is very advantageous for a gravity assist maneuver, since the $\theta_p$ value is not known a priori. 

The procedure used for the gravity assist is then identical to the one used for the orbit transfer in Sect. \ref{sec:num_transfer}.
The only difference is the quantity to minimize.
In the same way, the computational efficiency performance is identical to the one presented in Sect. \ref{sec:comp}.

\subsection{Results}

\begin{figure}
    \centering
    \includegraphics[width=0.58\linewidth]{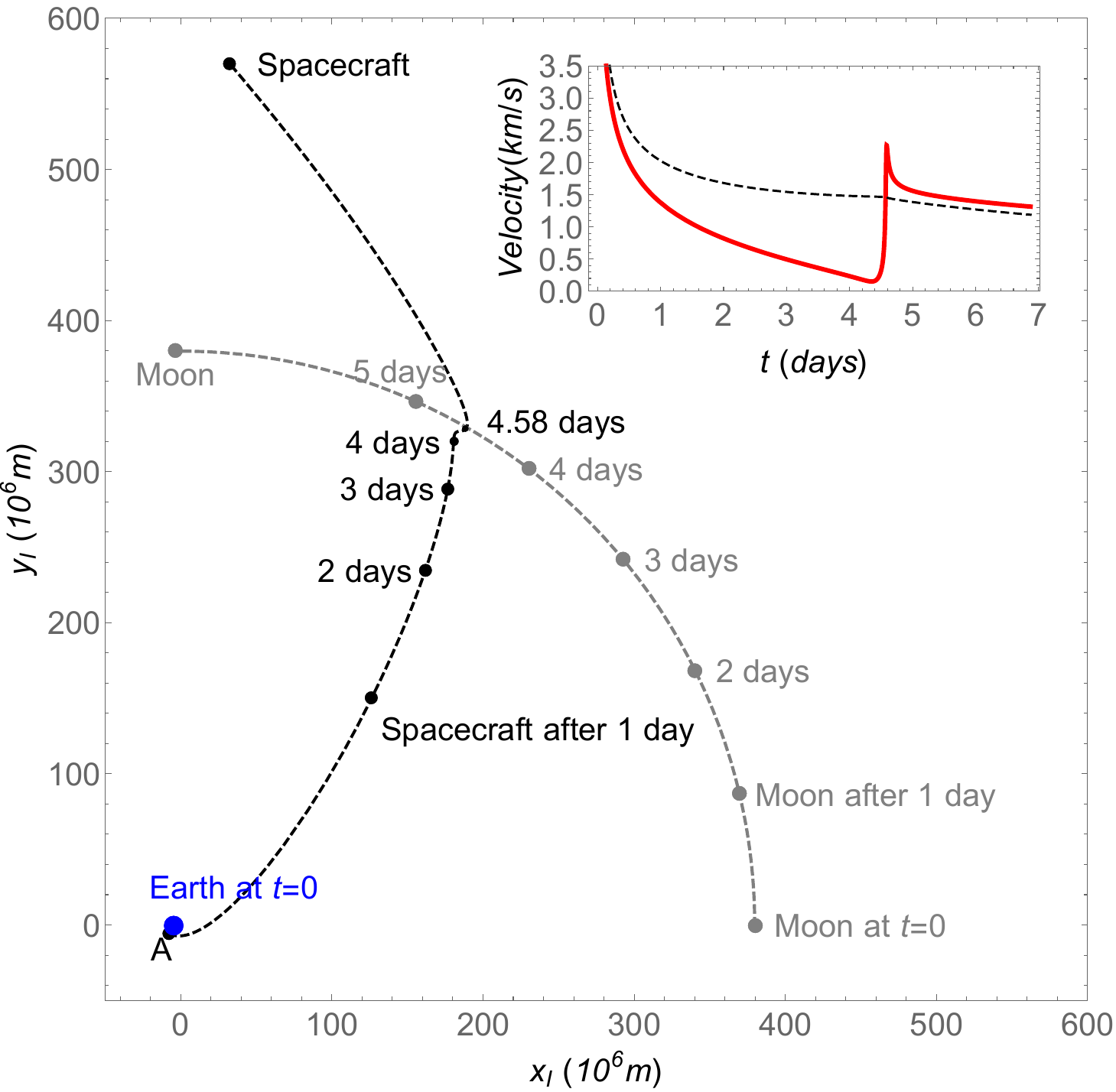}
    \caption{Trajectory obtained with the TV type constraints in TFC for a flight time $T=4.58\,\mathrm{days}$ and a periapsis altitude of $a_p=100\,\mathrm{km}$ in the inertial frame of reference associated with the Earth-Moon barycenter. In the inset, the red curve represents the magnitude of the velocity of the spacecraft, while the dashed black curve is the escape velocity from the Earth-Moon system.
    As the motion of the Earth is very close to the Earth-Moon barycenter, only the initial position of the Earth is indicated.}
    \label{fig:traj_GA}
\end{figure}

\begin{table}
    \centering
    \begin{tabular}{|r|c|c|}
    \hline
    $r_p\,(\mathrm{km})$ & $\Delta V\, (\mathrm{km/s})$ & $T\,(\mathrm{day})$ \\
    \hline
    50    &3134.6159	&4.58\\
    100   &3134.5947	&4.58\\
    150   &3134.5736	&4.57\\
    200   &3134.5527	&4.57\\
    500   &3134.4310	&4.57\\
    1000  &3134.2389	&4.56\\
    2000  &3133.8823	&4.55\\
    5000  &3132.9201	&4.51\\
    10000 &3131.4447	&4.45\\
    \hline
    \end{tabular}
    \caption{Minimum value of the velocity impulse $\Delta V$ required and corresponding flight time for each value of the periapsis altitude.}
    \label{tab:DV_T}
\end{table}

\begin{table}
    \centering
    \begin{tabular}{|r|c|c|}
    \hline
    $r_p\,(\mathrm{km})$ & $\DE\,(\mathrm{km/s})^2$ & $T\,(\mathrm{day})$ \\
    \hline
50      &1.6717 &   2.05	 \\
100     &1.6487 &   2.07     \\
150     &1.6266	&   2.09     \\
200     &1.6053	&   2.11	 \\
500     &1.4909	&   2.22     \\
1000    &1.3389	&   2.37     \\
2000    &1.1199	&   2.57	 \\
5000    &0.7553	&   2.82	 \\
10000   &0.4843	&   2.90	 \\
    \hline
    \end{tabular}
    \caption{Maximum value of the energy gain $\DE$ obtained and flight time corresponding for each value of the periapsis altitude.}
    \label{tab:DE_T}
\end{table}

\begin{figure*}
	\centering
    \includegraphics[width=0.7\linewidth]{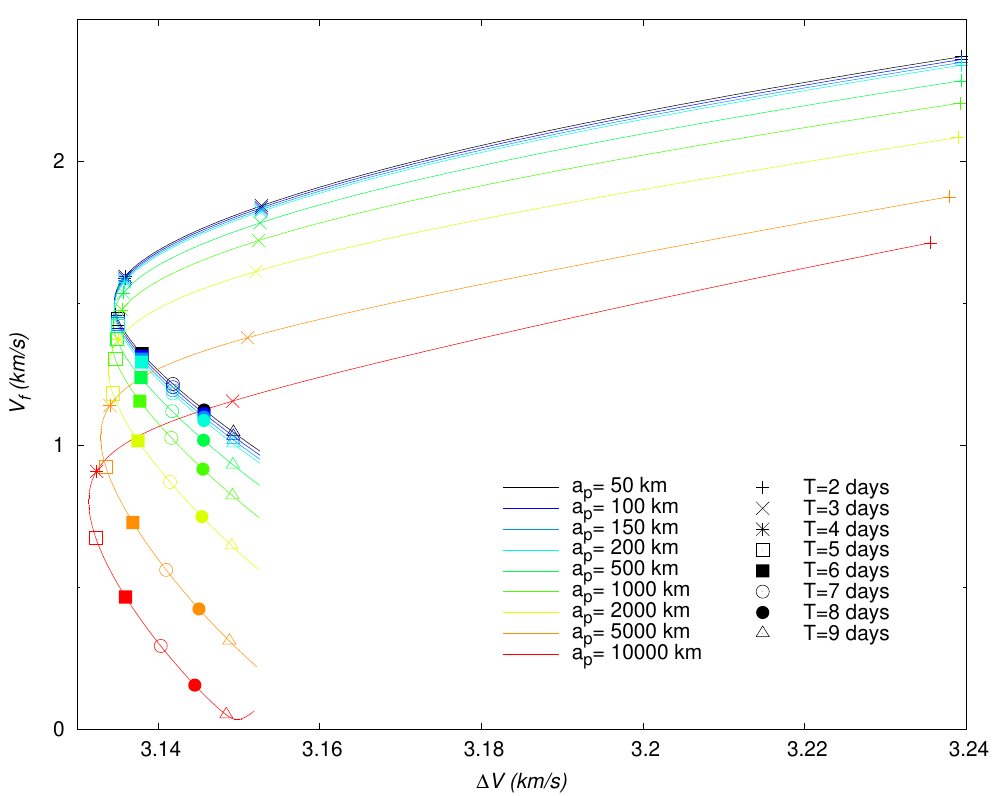}
	\caption{Velocity $V_f$ after the gravity assist maneuver with respect to the initial impulse $\Delta V$, and for different periapsis altitudes $a_p$. For each value of $a_p$, the curve is obtained with different values of the flight time $T$. The integer values of $T$ in days are indicated for each curve with a special symbol.}
	\label{fig:rend3}
\end{figure*}
\begin{figure*}
	\centering
    \includegraphics[width=0.6\linewidth]{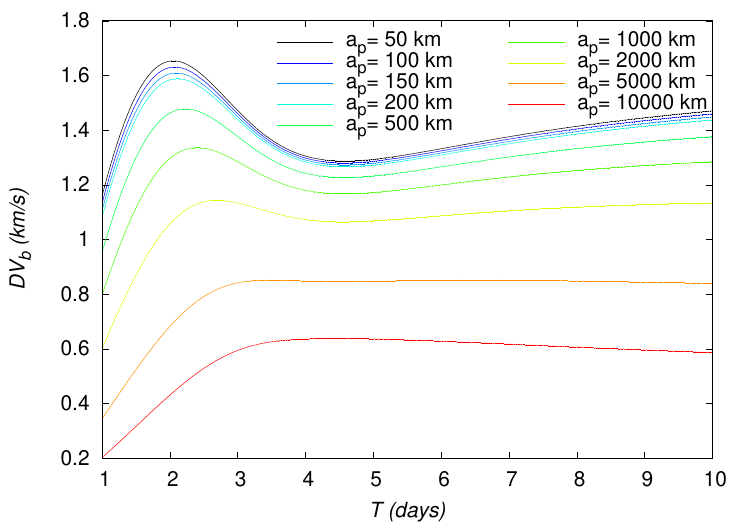}
	\caption{Velocity gain $\DVb$ of a rendezvous maneuver around the Moon, evaluated using the TV type constraints as a function of time for different periapsis altitudes $a_p$. The initial point is at an altitude of 167 km around the Earth.}
	\label{fig:rend1}
\end{figure*}
\begin{figure*}
	\centering
    \includegraphics[width=0.6\linewidth]{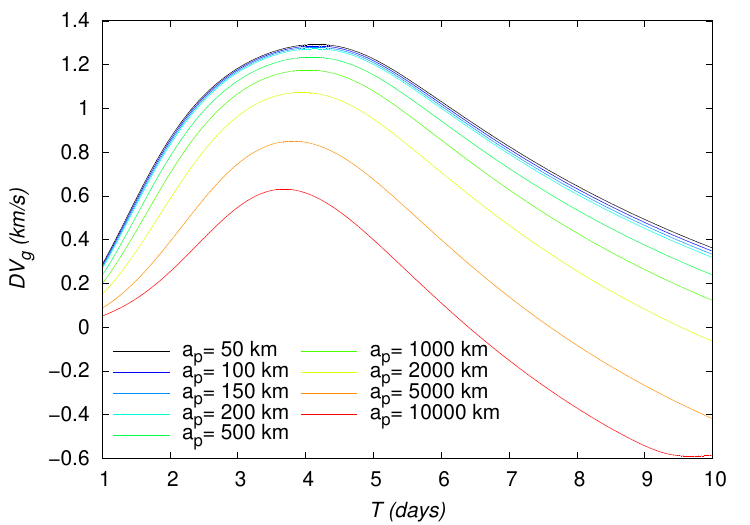}
	\caption{Velocity gain $\DVg$ of a rendezvous maneuver around the Moon, evaluated using the TV type constraints as a function of time for different periapsis altitudes $a_p$. The initial point is at an altitude of 167 km around the Earth.}
	\label{fig:DVg}
\end{figure*}
\begin{figure*}
	\centering
    \includegraphics[width=0.6\linewidth]{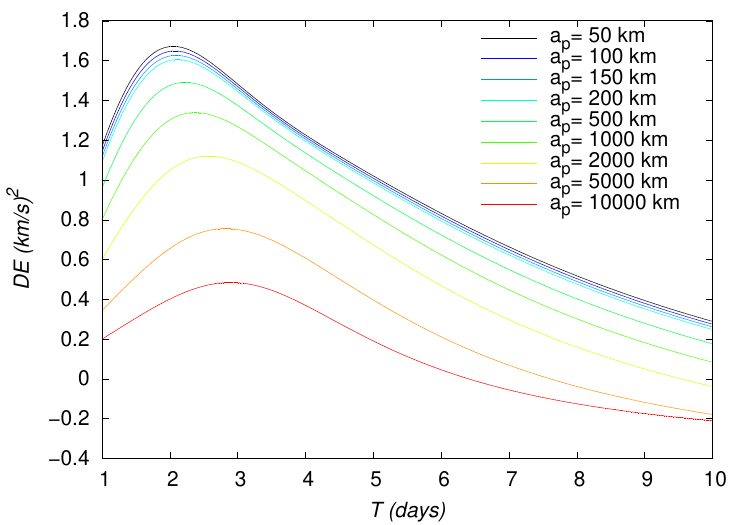}
	\caption{Energy gain $\DE$ of a rendezvous maneuver around the Moon, evaluated using the TV type constraints as a function of time for different periapsis altitudes $a_p$. The initial point is at an altitude of 167 km around the Earth.}
	\label{fig:rend2}
\end{figure*}
\begin{figure}
    \centering
    \includegraphics[width=0.6\linewidth]{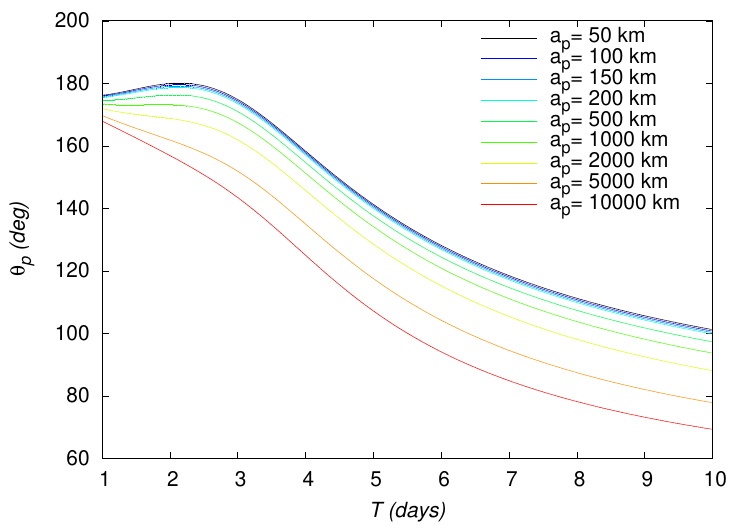}
    \caption{Periapsis angle $\theta_p$ of a rendezvous maneuver around the Moon, evaluated using the TV type constraints as a function of time for different periapsis altitudes $a_p$. The initial point is at an altitude of 167 km around the Earth.}
    \label{fig:thetaf}
\end{figure}
\begin{figure}
    \centering
    \includegraphics[width=0.7\linewidth]{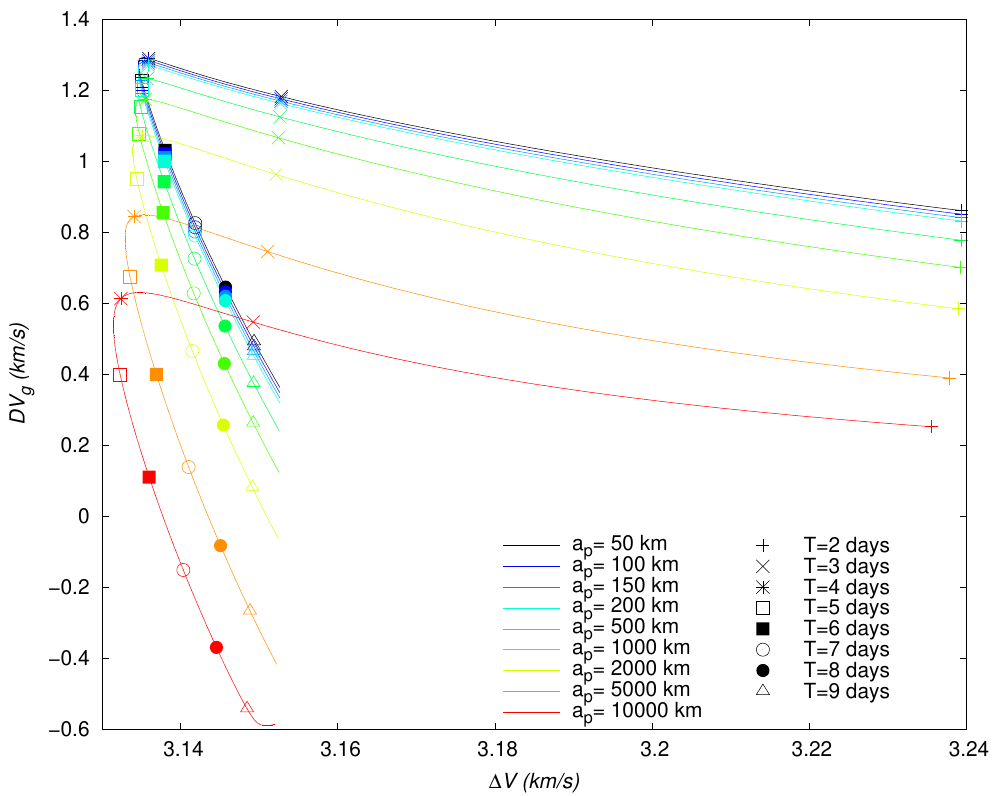}
    \caption{Velocity gain $\DVg$ after the gravity assist maneuver with respect to the initial impulse $\Delta V$, and for different periapsis altitudes $a_p$. For each value of $a_p$, the curve is obtained with different values of the flight time $T$. The integer values of $T$ in days are indicated for each curve with a special symbol.}
    \label{fig:DVg_DV}
\end{figure}

As was explained in Sect. \ref{sec:advan_GA}, one of the advantages of the TV type constraints is the ability to choose the periapsis distance $r_p$.
This makes it easier to observe its influence on the gravity assist since it is not deduced a posteriori from the solution.
For clarity, a new variable related to the altitude of the perigee is defined as $a_p=r_p-r_m$, where $r_m=1738~\text{km}$ is the radius of the Moon.
The procedure explained in the last subsection is performed for several values of the altitude of the perigee, given by $a_p=50$, $100$, $150$, $200$, $500$, $1000$, $2000$, $5000$, $10000$ km, imposed by the constraints for which the flight time varies along $1\leq T \leq 9.99$ days with a step of $0.01$ day.
An example of trajectory obtained with this procedure is shown in Fig. \ref{fig:traj_GA} for $T=4.58\,\mathrm{days}$ and $a_p=100\,\mathrm{km}$ using the frame of reference centered at the barycenter. The velocity with respect to the inertial frame is in red, and the dashed black line represents the escape velocity from the Earth-Moon system defined as $\sqrt{2(\mu_e+\mu_m)/r_b}$, where $r_b$ is the distance between the spacecraft and the barycenter.

Therefore, for each value of $a_p$, a curve representing the evolution of $V_f$ with respect to $\Delta V$ is obtained from the different values of $T$ in Fig. \ref{fig:rend3}.
From Fig. \ref{fig:rend3}, it can be deduced the specific impulse $\Delta V$ necessary for a given flight time in order to obtain a requested velocity $V_f$ after the gravity assist.
As expected, as the periapsis altitude $a_p$ decreases, the velocity increases.
The largest values of the velocity $V_f$ are obtained for short flight times ($T\leq 3\,\textrm{days}$), but they require a higher impulse $\Delta V$.
For longer flight times ($T> 3\,\textrm{days}$), a given value of $\Delta V$, and, consequently, a given quantity of propellant can lead to two different values of $V_f$, the highest one being given by the shortest flight time among the two possibilities.
This leads to the presence of a minimum value of $\Delta V$ and a corresponding flight time $T$ for each value of the periapsis altitude, whose values are given in Table \ref{tab:DV_T}.
Depending on the periapsis altitude, the minimum $\Delta V$ varies between $3131.4447$ and $3134.6159$ km/s, and the corresponding $T$ between $4.45$ and $4.58$ days.

The velocity gains, $\DVb$, defined in Eq.~(\ref{eq:dvb}), and $\DVg$, defined in Eq.~(\ref{eq:dvg}), and the energy gain, $\DE$, defined in Eq.~(\ref{eq:dener}), are also shown in Figs. \ref{fig:rend1}, \ref{fig:DVg}, \ref{fig:rend2}, for each value of $a_p$, respectively.
From Figs. \ref{fig:DVg} and  \ref{fig:rend2}, it can be verified that the gravity assist allows the spacecraft to gain energy and then to increase its velocity.
The evolution of $\DE$ presents a maximum different for each periapsis altitude $a_p$, whose value and corresponding flight time $T$ are indicated in Table \ref{tab:DE_T}.
With respect to $a_p$, the maximum $\DE$ varies between $0.4843$ and $1.6717\,(\mathrm{km/s})^2$,  for a corresponding $T$ between $2.05$ and $2.90$ days.
However, it can be observed that, for higher values of $T$, $\DE$ can be negative for $a_p\geq2000\,\mathrm{km}$. These negative values $\DE$ should correspond to the observed negative values of $\DVg$ and then lead to a decrease in velocity.
Nevertheless, this decrease in velocity cannot be indicated by the velocity gain, $\DVb$, which only enables to observe the absolute variation of velocity.
On the other hand, the peak of the velocity gain, $\DVb$, matches with a reasonable approximation with the peak of the energy gain, $\DE$, for $a_p\leq 2000~\text{km}$.
In Fig. \ref{fig:thetaf}, the evolution of the angle at the periapsis $\theta_p$ is represented with respect to the flight time.
The maximum gain in energy is for $\theta_p=180\,\mathrm{deg}$, according to Eq.~(\ref{eq:dener}), which coincides with flight times of about 2.2 days and values of $a_p$ up to 200 km.
The angle can be below $90\,\mathrm{deg}$ for large flight times if $a_p\geq2000\,\mathrm{km}$.
This was expected because, when the gravity assist maneuver leads to a decrease in energy, as this is the case for these conditions (Fig. \ref{fig:rend2}), $\theta_p$ should be below $90\,\mathrm{deg}$.
In Fig. \ref{fig:DVg_DV}, the velocity gain $\DVg$ is represented as a function of the initial impulse $\Delta V$.
It can be noted that $\DVg$ presents a maximum around the flight time of $4$ days, whose exact value depends on the periapsis altitude. 
It can then be observed that, for values of the velocity impulse close to the minimum (see Table \ref{tab:DV_T}), it is possible to slow down the spacecraft at the minimum or to speed it up at the maximum, depending on the objectives of the mission, by choosing suitable values for the flight time and periapsis altitude.
Therefore, the minimum quantity of propellant with only a small variation can be used for all the possible gains in velocity.



\section{Conclusions}
\label{sec:conclusion}

This paper proposed the use of the tangential velocity (TV) type of constraints in the Theory of Functional Connections (TFC) in order to find possible trajectories of orbital maneuvers.
The TFC method enables to include linear constraints, as initial or final conditions into a functional interpolation of the solution.
Therefore, any solution given by TFC will always analytically satisfy the constraints.
In the example of an orbit transfer from a circular orbit around the Earth to a circular orbit around the Moon, the constraints given to TFC in the case of the Two Point Boundary Value Problem (TPBVP) are the initial and final positions.
However, for such a problem, the choice of a tangential impulse enables to minimize the cost.
Therefore, imposing a tangential velocity constraint in TFC enables a direct search for solution solutions minimizing the transfer cost.
In this paper, it was shown that this can be only done in polar coordinates in order to keep the constraints linear.

As a result, using the TV type constraints in TFC significantly reduces the computational time.
Indeed, for the example of the orbit transfer from the Earth to the Moon, if the final velocity is imposed as tangential, only one optimization procedure is necessary to find the best position of the spacecraft on its initial orbit, as the final position is directly given by the solution.
In the TPBVP, two optimization procedures are necessary, one for the initial position, and another for the final one.
Using the TV type constraints in TFC also avoids the use of patched conics approximation and integrators like Runge-Kutta required by shooting methods.
Finally, this also avoids the risk of collision with the final body, because the final position given by the TFC solution corresponds to the periapsis of the trajectory around the final body, thus making it possible to choose a small radius for the final position.

The use of the TV type constraints in TFC is applied in this paper in the cislunar space for two maneuvers: an orbit transfer from a circular orbit around the Earth to a circular orbit around the Moon, and a lunar gravity assist maneuver.
For the orbit transfer, where the objective is to find the trajectory minimizing the quantity of propellant required, it has been shown that the TV type constraints yield identical results to those of the TPBVP with a significant decrease in computational time.
Concerning the lunar gravity assist, the TV type constraints enable a rapid determination of the minimum initial impulse with respect to the required final velocity (after the encounter) depending on the different parameters of the gravity assist.
Indeed, the TV type constraints enable to impose the altitude of the periapsis of the gravity assist, which is one of the determinant parameters, being the two other parameters given by the solution.
It can be noted that, for both applications considered here, the computation performed with TFC is identical, and that the unique difference lies in the quantities to minimize.

The choice of the cislunar space was only made in order to work in a well-known and studied framework, where the obtained results can be compared with previous reference works. However, the TV type constraints are not limited to the cislunar space, but could be used in any spatial environment.
In the same way, the use of the TV type constraints is not restricted to the two maneuvers considered here, orbit transfer and gravity assist, but could be applied in any spatial environment or maneuver requiring a tangential velocity constraint.
As TFC turned out recently in several studies as an effective and promising method to design orbital maneuvers, and as a tangential velocity in orbital maneuver corresponds to a optimal solution related to the lowest propellant consumption, the use of the TV type constraints in TFC could then have many other useful applications in the future. 







 





\bibliography{ref2}



\section*{Acknowledgements}

The team acknowledges further support from ENGAGE-SKA Research
Infrastructure, ref. POCI-01-0145-FEDER-022217, funded by COMPETE 2020
and FCT, Portugal; IT team members acknowledge support from Projecto
Lab. Associado UID/EEA/50008/2019. We acknowledge  support by the
European Commission H2020 Programme under the grant agreement 2-3SST2018-20, and 
support from Center for Mechanical
and Aerospace Science and Technologies - C-MAST, funded by
FCT-Fundação para a Ciência e a Tecnologia, Portugal through projects
https://doi.org/10.54499/UIDB/00151/2020 and https://doi.org/10.
54499/UIDP/00151/2020. This work is also supported by CFisUC (UIDB/04564/2020 and UIDP/04564/2020). This work is supported by the Fundação para a Ciência e Tecnologia (FCT), Ph.D. grant No.2022.12341.BDANA. DM, AKAJ and TV were partially supported by the project Centro de
Investiga\c{c}\~ao em Ci\^encias
Geo-Espaciais, reference UIDB/00190/2020, funded by COMPETE 2020 and
FCT, Portugal. AKAJ and TV acknowledge partial support from ATLAR internal
project n. 01012024. VMO is partially supported by the São Paulo Research Foundation (FAPESP, Brazil), under Grants No. 2021/11306-0 and 2022/12785-1.

\section*{Author contributions statement}

AKAJ and TV conceived the idea, generated the figures, and wrote the first version of the manuscript. All authors analyzed the results and reviewed the manuscript. 

\section*{Additional information}

The authors declare no competing interests.



\end{document}